# Simulation Typology and Termination Risks

*Alexey Turchin*
Foundation Science for Life Extension
alexeiturchin@gmail.com, corresponding author

*Michael Batin*
Foundation Science for Life Extension

*David Denkenberger*
Alliance to Feed the Earth in Disasters
University of Alaska Fairbanks
ddenkenberger@alaska.edu

*Roman Yampolskiy*
University of Louisville
roman.yampolskiy@louisville.edu

## Abstract

The goal of the article is to explore what is the most probable type of simulation in which humanity lives (if any) and how this affects simulation termination risks. We firstly explore the question of what kind of simulation in which humanity is most likely located based on pure theoretical reasoning. We suggest a new patch to the classical simulation argument, showing that we are likely simulated not by our own descendants, but by alien civilizations. Based on this, we provide classification of different possible simulations and we find that simpler, less expensive and one-person-centered simulations, resurrectional simulations, or simulations of the first artificial general intelligence's (AGI's) origin (singularity simulations) should dominate. Also, simulations which simulate the 21st century and global catastrophic risks are probable. We then explore whether the simulation could collapse or be terminated. Most simulations must be terminated after they model the singularity or after they model a global catastrophe before the singularity. Undeniably observed glitches, but not philosophical speculations could result in simulation termination. The simulation could collapse if it is overwhelmed by glitches. The Doomsday Argument in simulations implies termination soon. We conclude that all types of the most probable simulations except resurrectional simulations are prone to termination risks in a relatively short time frame of hundreds of years or less from now.

*Nomenclature*:
AGI – Artificial general intelligence
BB – Boltzmann brains
DA – Doomsday argument
ETI – Extraterrestrial intelligence
NPC – Non-playing characters
OM – Observer-moments
P-zombies – philosophical zombies (equivalent to NPC)
SA – Simulation argument
UFAI – Unfriendly (superintelligent) AI, Evil AI

**Introduction**

Bostrom's famous article (Bostrom, 2003) and the movie trilogy "Matrix" introduced our culture to the idea that we may live in a simulated world, created by some sophisticated artificial general intelligence (AGI). They also provided a powerful explanatory narrative of "glitches", which are supposed to be bugs in the performance of the simulation. These are observed as mundane strange events, which are different from the "paranormal" type of strangeness.

Living in a simulation could have some observational consequences: non-typicality of an observer's position in the world, glitches, miracles, computer viruses, cheat-codes (some form instruments to control simulation from inside which bypass its physical laws), direct communication with simulation owners and the simulation's termination. Whitworth argues that even the character of the physical laws of our universe (e.g. the speed of light) are best explained by the limitations of computing processing on the base reality level (Whitworth, 2008).

The problem of simulation termination was listed in Bostrom's classification of existential risks or X risks (those risks that could cause human extinction or a permanent reduction in the potential of humanity) as one of the most important (Bostrom, 2002), but it remains underexplored. However, interest in the topic is growing. Hanson wrote that we should live the most interesting lives we can in order to make simulation owners be interested in simulation continuity (Hanson, 2001). Torres explored risks of multilevel simulation (the simulation runs a simulation) termination (Torres, 2014).

Greene wrote in 2018 "Termination Risks of Simulation Science" (Greene, 2018) about "decision-theoretical uncertainty": If someone starts a simulation of her past but commits to terminate the simulation if it becomes nested (the simulation starts a simulation), she basically has given herself the death penalty, as she mostly likely already in such a simulation. The second termination risk Greene explored is testing whether we are in a simulation or not. At the time of this writing, two physical experiments were performed with such a goal: 1) scattering of cosmic rays on the possible space-time lattice of the simulation (Beane, Davoudi, & Savage, 2014) and 2) the test of accumulation of soft errors in computer memory on the Earth's surface and below the ground, which demonstrated that such soft errors appear in our universe, but not in the "base reality" (Alexander, 2018). A rate of soft-errors in hardware – bit flips caused by interaction with cosmic ray radiation – was measured. If we were in simulation but errors will be caused by "real" radiation, such rate will not depend on the location of the computer. However, after the computer was placed in the radiation protected underground bunker, the rate of errors significantly declined, which is the evidence that these errors are part of our world, not of the simulator's base reality.

Greene argues that the utility of such experiments is negative as they either create negative results, in most cases, or they confirm that humans are in a simulation, but this makes the simulation useless for their owners as a historical reconstruction and then the simulation could be terminated.

To solve the question of possible observational consequences of living in the matrix (and associated catastrophic risks), we firstly in Part 1 patch the simulation argument (SA), explore a possible simulation typology and estimate what type of simulation we most likely live in. In Part 2 we analyze the risks of simulation termination.

**Part 1. Simulation**

*1.1.     A patch to the Simulation Argument*

*1.1.1. Bostrom's simulation argument*

In Bostrom's classical simulation argument, it is claimed that at least one of three alternatives is true (Bostrom, 2003):

1) We are living in a computer simulation.
2) We will become extinct before the creation of a powerful AI capable of creating detailed simulations.
3) Future civilization will be completely not interested in the creation of ancestor simulations.

This is based on the observation of rapidly expanding computational ability that could allow an advanced civilization to produce many ancestor simulations. This argument is in some sense similar to the classical Doomsday Argument (DA), again explored by Bostrom: it assumes the existence of only human civilization and then tries to deduce our location in time in this civilization history (Bostrom, 2013; Gott, 1999). If one goes by number of humans, humanity now is likely to be near the middle of cumulative humans, which means humanity may only have a few centuries left to exist if populations remain near their current level.

*1.1.2. The circularity problem of the simulation argument does not disprove it*

In the case of the SA, it results in some form of circularity as was mentioned by critics: if one assume that humanity is in the simulation, one cannot guess anything about our creators by definition, including whether we are ancestors of the simulation creators (Medvedev, 2003). This does not appear to be a significant criticism, because in this objection it is already assumed that we are in a simulation. However, it puts into question our guesses about the power of computers or ethics of simulation creators as well as whether they are human descendants at all.

*1.1.3. Universal Simulation Argument: alien simulators also should be taken into account*

There is another version of the DA, called the "Universal DA" in which it is assumed that our civilization is typical of all possible civilizations and thus its age could be used to guess the age of a typical civilization – including ours (Gerig, Olum, & Vilenkin, 2013). In the same way, a Universal SA could be suggested:

One of the alternatives is true:

1) We are living in a simulation.
2) There are no civilizations, AIs or any non-agential processes which create simulations in the whole universe (or multiverse (Tegmark, 2009) if it actually exists) or they are extremely rare.

The main idea is that the simulation of human civilization could be created by a completely non-human creator. This is analogous to humans creating computer games such as StarCraft full of magical dragons, but our civilization does not include magical dragons. This is a significant criticism of the third alternative from the original SA: that our descendants will choose not to create simulations based on ethical reasons. If the universe is infinite, many different approaches to simulations will be embodied.

The Universal SA also makes much weaker the second leg of the Bostrom's SA, that "we will become extinct before the creation of a powerful AI capable of creating detailed simulations," as inevitable humanity extinction doesn't prevent other civilizations from simulating us. However, if humanity's extinction is inevitable, it is evidence for the fragility of all other possible civilizations, as we are typical and inevitable extinction requires very universal mechanism. Extraterrestrials could come to Earth in the remote future after human extinction, find our DNA (Turchin & Denkenberger, 2019; Yampolskiy, 2018) and recreate us, perhaps in a simulation. Or it could be that Earth never existed in reality, but a completely non-human intelligence is investigating different combinations of initial parameters of biological evolution.

There is also a clear reason why an extraterrestrial intelligence (ETI) would simulate humanity: any ETI needs to solve the Fermi paradox (why a civilization does not observe alien civilizations). Running many simulations of other possible civilization could provide an estimation of their probability distribution in the universe (Dainton, 2002; Turchin & Denkenberger, 2019). This could be called this a Fermi Simulation.

In the Universal SA, the idea that simulations are impossible is much weaker than in Bostrom's SA, so it is much stronger evidence for alternative (1): we are in a simulation. However, Bostrom bases the evidence of his Simulation argument on the trends of computer development on Earth, which is not required for the Universal SA, which is based on general power law that more simple objects are more numerical, as we will discuss in next subsection.

*1.1.4. Occam's Razor for simulations: illusionary simple worlds dominate over real and complex ones*

To argue for Universal SA, we suggest the following claim:

Occam's Razor for simulations: For any observation X, there are many worlds which could create it, and between them dominate those worlds which create X using the simplest – or cheapest – computations (basically, we apply here to Solomonoff's induction based on Kolmogorov complexity and we get something like an Occam's razor for observations, and this is not new (Hutter, 2000)) Yampolskiy posed a similar generalized sample attribution problem and suggested, using the same logic as in SA, that most of life in the Universe has synthetic origin (Yampolskiy, 2016).

Now we will define the simulation of the world W: a world W' is the simulation of the world W, if it provides the same observations as the world W using fewer computations than the intelligent observer expects based on his/her internal model of W. (Note that this definition does not require a simulator or an AGI.)

From the definition of simulation and Occam's razor for simulations, it follows that simulations should dominate over real worlds (as real things are typically much more computationally complex than their illusions.)

This could be illustrated in real life: in most cases, when one sees a rare object or occurrence, one, in fact, sees only its simulation. For example, one much more often sees nuclear explosions in movies/videos than in real life (never for most people). One more often sees photographs of beautiful houses than in real life. Photographs, dreams, YouTube videos, computer games – all these are (partial – not whole world) simulations of some real or hypothetical things. However, in the real world, some observations are still dominated by real things – like seeing the sun. But the more time people spend on computers, the more often they see photos of the sun instead of the real one. In other words, the lower price of computation increases the proportion of illusions to real things.

Next, we define a computer simulation: a simulation where a computer is used to effectively mimic the observations of a real object or occurrence. The use of a computer to mimic observations is very effective in a computational sense, as a computer "knows" how to adjust the observation so it will approximate the expected object in the most effective ways. For example, a computer game only creates textures of walls which are observed and does not calculate unobserved parts.

To get to Bostrom's SA from this point, we should add the expectation of the abundance of computing power available for advanced superhuman civilizations in the universe. This cannot be directly derived from Moore's law, as this law is just an observation from Earth, which could be a simulation. However, the idea of technological evolution which lowers the price of computation is Earth independent. In other words, if at least some share of civilizations reaches high technological level, at least part of them could create enormous amounts of simulations of different civilizations. Such a civilization running simulations may exist in a Universe with completely different physics than ours.

*1.1.5. Modal realism obstacle*

Also, the Universal SA, which looks over a potentially infinite universe, has a "modal realism obstacle": If, as it is assumed by the modal realism (Bostrom, 2002b), everything possible exists, then there are an infinite number of real worlds and an infinite number of simulated worlds. Infinities are notoriously difficult to compare. One could escape infinities by calculating only the number of simulations in the observable universe (or some analogue if our simulation is misleading us about the physics of the base reality).

*1.1.6. SA objection: Could the simulation hypothesis be falsified?*

This objection runs as follows: the hypothesis that humanity is in a simulation cannot be falsified if one assumes that the simulation is perfect. However, if the simulation is not perfect, it is not a simulation.

Such perfect simulations do not have glitches or miracles, and there is no way to escape them by definition. Thus, the claim that one is located in such simulation does not have any "meaning".

However, this objection is only semantics, as even if the simulation does not have any observable evidence that it actually is a simulation now, the simulation could get such this evidence any time its owners want to create it – or terminate it.

Moreover, any simulation is computationally constrained by computing power available to its owners and thus it will be terminated as soon as the task for which the simulation was created is fulfilled. This termination will likely occur much earlier than the "natural end" of the universe by heat death. There are some unconventional computing methods like reversible computations, potentially allowing infinitely long simulations running on finite amount of energy (Bennett, 2003). However, there could be other reasons why the simulation run-time will be limited such as limited attention span of the hosts or memory limits to preserve the simulation's results.

*1.1.7. SA objection: "location" term is only applicable to material objects, but not computational processes, and is meaningless in the case of modal realism*

If we dig deeper into the question of "location", we should note that a mind – as a computational process – may be located simultaneously in all places where the given computational process is performed. For example, the number 12 is located everywhere where 12 objects are placed, no matter if they are 12 apples, 12 oranges or 12 matches.

If modal realism is true, and everything possible exist, then "being in some place" loses any definite meaning, as the same mind is located in all possible places which do not contradict its existence.

The thesis of "being in a simulation" becomes meaningful only if we use it to estimate the probability of glitches or other evidence of hosts' activity. If it cannot be falsified, it becomes pure speculation, which is similar to speculations about the "correct interpretation" of quantum mechanics (Tegmark, 1998).

*1.1.8. SA objection: the measure is proportional to the energy of computations, and a computer simulation will have a much lower measure than reality*

Another argument against Bostrom SA is "measure inequality" between real minds and simulations. If there are two minds, one biological, and another is its exact computer simulation, an observer could assume that she has equal probability to be any of them. Bostrom escapes the measure problem by postulating the "bland indifference principle": "if there are many more simulated realities than non-simulated ones (and the two are indistinguishable), then we are probably in a simulated reality" (Bostrom, 2003).

However, this assumption could be challenged, as any computer mind could be split (at least theoretically) into two parallel working minds, which use half as much energy each. See more about splitting minds in (Yudkowsky, 2008) thought experiment with "Ebborians". In this thought experiment Ebborians live as simulated minds in a two-dimensional world, which have some thickness in the third dimension. Such two-dimensional mind could be sliced parallelly into two minds with equal functionality, and so on until there will be many copies of exactly the same mind. This creates a problem in calculating the actual number of the copies and thus in calculating the probability of being one of the copies. Actual physical computers also could be viewed as Ebborian computers as every piece of data is presented by an electric charge, which could be theoretically split in two.

Obviously, such splitting of computations cannot be performed infinitely: there is something like "Plank level of computations": minimum energy required to perform given computations. It is natural to suggest that the "measure of existence" is somehow related to the number of "Plank computations" in a given implementation of a mind. This conjecture could be proved by contradiction: if the measure of existence is not dependent on the number of Plank computations, then our estimation of the measure will depend on the arbitrary and illusionary splitting of a mind into non-existence parts. Also, such dependence should be linear, or else one could get a counterfactual capability to manipulate measure by selecting different parts.

The number of the "Plank computations" is proportional to the total energy involved in the computation, so, in short, we just preliminary demonstrated that the measure of existence of a mind is equal to the energy of computations used for its simulation.

This result is important as it could be used as an argument against Boltzmann brains (BBs) of many types. These brains – being quantum fluctuations – become much more probable at less energetic levels than wet brains.

If the measure depends on the energy of calculation and simulations are very energy efficient, they will get less "measure mass" (see section 1.3). But if a large number of simulations exist, this could override lower measure of individual simulations. For example, if each simulation is million times more energy efficient than the human brain, but there are 1 billion simulations, then a random observer will still most likely find herself in a simulation.

To account for measure difference between simulations and real brains, we should assume a parameter of the energy efficiency of simulation X, which could be of several orders of magnitude higher than the human brain, especially if simulations are implemented via very cold superconductive computers in the remote future. However, as the human brain is fairly energy efficient, X is probably not prohibitively large. More discussion about energy requirements of human brain computation can be found in Sandberg's article (Sandberg, 2016). Also, less advanced future AIs will use more energy for human brain simulations, and more advanced could use less energy. But this does not mean necessarily that our simulation is created by simpler AGI, as more advanced AGI may create a much larger number of simulations, even tile the whole galaxy[1] with them. Alternatively, more advanced AGI could deliberately create a simulation which consumes more energy, in order to get measure dominance and thus control over most observers. This could be done for the resurrection of the dead or for preventing human suffering created by evil AGIs (Turchin, 2019).

*1.1.9. SA objection: Domination of the simulations without consciousness*

There is also the consciousness inequality counterargument: If simulating phenomenological consciousness is computationally expensive, but practically useless, then simulations may be dominated by philosophical zombies (p-zombies) experiencing no consciousness (here we are leaving aside the question if p-zombies are possible at all) (Chalmers, 1996). In this case, having real phenomenological consciousness is an argument for being in the real world. This would be true at least for the cheapest and most abundant simulations (see Section XX). However, consciousness can be a general side effect of computation (Yampolskiy, 2017).

*1.1.11. SA objection: If the simulation of conscious minds is possible, random mathematical minds should dominate*

In the post on the blog LessWrong (TheTripleAffirmative, 2019) it was suggested that if digital simulating of conscious minds is possible, then "parasitic" host-less simulations – appearing from random processes in nature or just as parts of the mathematical world – should numerically dominate over really existing minds and real computer simulations.

The weak point of the argument is that one cannot prove that one is not a random chaotic mind, and moreover, even random observer-moments could form "chains", as was describe by Egan in his dust theory (Egan, 2009), Wei Dai's UDASSA (Christiano, 2011) and by Mueller's article "Law without law" (Mueller, 2017).

In other words, humanity could be now inside a simulation without a creator, just a random book in the Babel library (an imaginary library from a Borges' novel (Borges, 2000), which consists of all possible books). There is no way humanity could prove that it is not in the simulation: the world humanity observes could be completely random, but as humans' thinking processes are also random, so humanity could conclude that the world is not random in 50 percent of cases – because any conclusion "yes" or "no" is purely random and is not based on any logic for BBs.

*1.1.12. SA for night dreaming*

For a better understanding of the logic of the SA, one could test it in a more familiar case, where people are in a simulation: night dreaming. Most people have night dreams but are not capable of recognizing them as dreams; however, a person could be trained to become lucid in dreams (Gackenbach & LaBarge, 2012). Such training mainly consists of learning the skill of detecting glitches in the dream and correctly interpreting them as signs of dreaming. Such glitches include all types of inconsistencies: changing objects, seeing dead people, jumping over significant stretches of time, not being able to read well, superhuman capabilities. Dreams are full of such contradictions, but as critical thinking in dreams is suppressed, it is very difficult to recognize them, and thus lucidity in dreams is rare. (In this sense dreams are similar to BBs, which are not capable of recognizing that they are BBs.)

---

[1] Though the galaxy could be a concept internal to our simulation, and we do not know if it is a general concept outside of it.

The SA for dreams runs as follows: the more unlikely event I observe, the more probable it is that I am dreaming about it. For example, if one wins $1 million in a lottery, this much more often happens in dreams than in reality, so one should think that it is likely that one is dreaming.

The real life of people is typically uneventful, but dreams are many times full of significant events, such as deaths, wars, accidents, and adventures. Thus, observing any significant life event could be evidence for dreaming.

SA for dreaming can be used to show the falseness of the main objection about the SA, suggested by Danila Medvedev: If one is in a simulation, one cannot make any conclusions about reality and thus cannot prove that one is in a simulation. (Medvedev, 2003) For example, one meets an apparently omniscient alien, and it tells one: "You are dreaming." Medvedev's argument for this case is the following: If one is dreaming, then one should not believe anything the alien said because one cannot trust information in dreams. However, there are two options: 1) the alien is real, and thus tells the truth, and one is dreaming, and 2) the alien is not real, thus one is dreaming about the alien. In both cases, one is dreaming.

*1.1.13. Unique self-location as an argument for simulation*

Besides possible glitches and miracles, or general considerations about numbers of real worlds, there is a third type of evidence for simulation: unique personal location in world history. Simulations are much more likely to simulate interesting people in the interesting moments of their lives, while in real life ordinary observer moments of ordinary people should dominate.

Thus, an observer should estimate the uniqueness of her/his position, but only include those observers which could be aware of the possibility of being in a simulation. In other words, the observer cannot be a $17^{th}$-century peasant, as s/he cannot know about the simulation. The correct question one should ask oneself is, "Am I a typical member of the class of all people who are interested in simulation?"

For example, one of the proponents of the simulation idea is Elon Musk (Wall, 2018), and he may think about the uniqueness of his position as evidence that he lives in one-person-centered simulation. (But the same information about him means little nothing for the rest of us.)

People also tend to overestimate the uniqueness of their position, as they tend to think about their in-group as special and about others as ordinary people [ref]. Any argument about the uniqueness of one's position not supported by any independent metric, e.g. extreme wealth, should be taken skeptically, as people are typically overconfident.

On the other hand, it could be argued that people working on global catastrophic risk prevention and mitigation are in unique positions that may participate in defining goals of the future AGI (Yampolskiy, 2019).

Such uniqueness of position should cause a surprise. As such a group of people who are working on X risks is probably on the order of magnitude of 1000 people, this could be taken as evidence of humanity being in a Fermi simulation, as only this people could be truly conscious as they are simulated in details, and other could be non-playing characters as they are simulated only when they are interacting this playing character. This type of simulation needs to simulate in detail only those people who could affect the outcome, and others could be non-playing characters.

However, "X-risk reduction group" is overlapping with the group of those who think about the SA, which makes the actual probability assessment difficult. One should only use those who know about the SA as the members of the reference class, from which an X risk reduction person is supposedly randomly chosen. In other words, only if the simulation-thinkers class is significantly larger than X-risk reducers class would it be surprising.

### *1.2. Typology of possible simulation as an instrument to find the most probable type of computer simulation in which we could be located*

The question which we will try to answer in this section is: "what is the most probable type of simulation humanity could be in?" We assume for now that the probability of being in a simulation is proportional to the number of simulation runs, and that being a physical human and a simulated mind have equal probabilities. (The alternative point of view where the measure depends on the energy of the computations was explored in subsection 1.1.8.)

*1.2.1. The most probable goal of a simulation*

Now we will list possible goals of creating detailed simulations by an advanced future civilization and will estimate how many simulations are needed for such goals.

1) AGI Safety research
1a) AGI confinement – virtual reality to box first AGI. The number of such simulation estimation: 10-1000s.

1b) Indexical uncertainty by confined AGI. Early unfriendly AGI simulates its owners to overcome them or create indexical uncertainty, that is a situation when a real mind has many copies in a simulation and thus can't know for sure if it is in the simulation or not. This could be used for blackmail. This was described in the post by (Stuart Armstrong, 2010). 10-1000s.

1c) Indirect normativity – Simulating humans for indirect normativity (the way where human values are not formally described, but the condition of their appearing are stated), suggested by Bostrom (Bostrom, 2014) and detailed by Christiano's post (Christiano, 2012). The number if such simulation will be probably just a few.

1d) Singularity simulation – Testing of different outcomes of the technological singularity, which can be defined as the advent of self-improving AGI (Good, 1965). This testing would different initial conditions. Future AGI will probably have an instrumental need to explore its origins. One reason for it is explore the distribution of different AGIs in the physical universe and is similar to Fermi simulation. Another reason is to explore the distribution of other AGIs in the universe, in order to establish acausal trade with those that cannot be contacted (Turchin, 2018b) or to identify the proportion of evil AGIs, which could cause astronomical suffering risks (s-risks) (Sotala & Gloor, 2017). In the case of evil AGI simulations, such simulations could be used to create indexical uncertainty in them, so they will be "afraid" to actually harm humans. This is a part of "Nelson's attack", where using humanity's earlier position in time, humans precommit to ask future Benevolent AI, if we ever create it, to simulate all potentially evil AGIs (Turchin, 2017). The number of simulations needs to be high to create indexical uncertainty in all possible evil AGIs, perhaps trillions. Self-improving AI also will have a need to test its new versions for safety and stability (Turchin & Denkenberger, 2017), and creation of singularity simulations may be a good way to find if new AIs design is safe.

1e) Data generation – If future AGI were still based on neural networks, but not on some first principles logic, it will need a lot of data for training. It that case, to create a new AGI or a new version of a given AGI as part of its self-improving process, a large simulation is need to generate a lot of training data. While the actual number of data-generating simulations is difficult to estimate, it probably should not be very large, as even a few simulations could provide a lot of data.

2) Scientific research
2a) Fermi paradox solutions. There would likely be many numerical solutions of the Fermi paradox by simulating different civilizations and different types of AGIs which they could create; this is useful for estimating X-risk probabilities (Turchin & Denkenberger, 2019).

2b) Problem-solving. Humans could be simulated to make them solve scientific problems. This is unlikely to be an effective way of the problem-solving for an AGI.

2c) Sociological research. The AGI may be curious about human history. Greene thinks that a large share of simulation will be devoted to historical experiments: running history with different initial conditions (Greene, 2018).

3) Resurrection simulations
Using incomplete data collected by digital immortality, a future superintelligent AI could resurrect the people who lived on the past by running a full simulation of the whole world (Tipler, 1997; Turchin & Chernyakov, 2018). If it is done correctly, it could be done just once, at least for any actually playing observer. Quantum randomness could be even helpful, as the whole setup will be partly similar to quantum random mind generator, which is capable to resurrect all possible minds in different branches of Everettian multiverse as was suggested by Almond (Almond, 2006). This is the opposite of non-playing characters (NPCs) from videogames which are barely simulated bots lacking their own consciousness, and which are simulated many times. Holes in the knowledge about past people could be filled with random quantum noise. If there were a quantum multiverse and acausal trading between branches, this could produce an exact resurrection of any person – in one of the branches. More in (Turchin, 2019).

4) Entertainment of high-level beings
Most of the currently created simulations – movies, games, books – are created for entertainment, and extrapolating this trend in the future suggests that most of the future simulations will be for entertainment. For example, Yampolskiy suggested "personal universes" as a solution to AGI safety problem and such personal universes for every human will be probably full of simulations of human history (Yampolskiy, 2019). However, most people's lives are likely too boring to be an interesting game from a simulator's point of view. Or the real player appears in the game rather rarely to act, and most of the time we are just semi-autonomous units.

5) Measure manipulation

Future advanced AGI could try to get control of almost all observers' experience in the world. This could be used as a protection against eternal suffering risks created by evil AGIs and ensure that most observers who have existed are happy ones. Such simulations are by definition made to increase the probability that an observer will find her/himself in it. However, its feasibility and desirability are questionable (Turchin, 2018). Such measure manipulation may be combined with a resurrectional simulation. If a future AGI wants to resurrect the dead, it may also want the biggest probability for any given human to be revived in a such resurrectional simulation, but not in some other experimental simulation where they could suffer.

From this, it is likely that the most probable simulations are the scientific ones (Fermi paradox solving and singularity modeling), resurrectional, measure domination or entertainment simulations.

*1.2.2. What observer's location tells the observer about a possible goal of one's simulation*

If a conscious observer lives in a simulation which is not simulating all minds of all people, but only some minds of "playing characters", then observer's own location – assuming that the observer is a playing character and able to recognize oneself as such – provides the observer information about a possible goal of his-her simulation.

For example, if an observer thinks that the observer lives in entertainment simulation, the observer should expect that his-her life should be interesting and important, and probably full of adventures. If one lives an adventurous life of the Count of Monte Christo, one should estimate higher the chances that s/he is in an entertainment simulation. If one lives a boring life of a medieval peasant, one is less likely to be in this (or any) type of simulation.

One could also be in a resurrectional simulation, but it should just mirror one's real life until the end. At the end, one would find oneself alive in a much more interesting world. The chances of being in the resurrectional simulation are higher if one has deliberately expressed interest in being in such resurrection or has collected digital immortality data (Volpicelli, 2016).

Many people think the singularity is near (Kurzweil, 2006). If one lives near singularity and is close to the circle of people who may define the shape of singularity by programing the first AGI, which could dominate all future history, this is evidence that one lives in a singularity simulation.

*1.2.3. The size of the simulation*

Simulations could be of different size in time, spatial extent and detail of accuracy. The size of the actually simulated region of space could be from the whole universe to only one observer's experiences. Special cases are probably simulations of the whole Earth, as it is what is needed for the experiments in human history, and of just one person's brain.

In the same vein, the time duration of a simulation could be from trillions of years to a fraction of a second, which special cases of the whole humanity history (5000 years) and one person' life.

The level of detail of the simulation could vary from full simulation of each atom to rough simulation of only observed experiences. The simulation could be even more coarse than in "reality".

There are probably two main type of simulations: physics level and observer-experience level. Even the most advanced high-fidelity computer simulation of the physical world can't predict the real world due to the quantum uncertainty and complexity. Some shortcuts may be needed to simplify the computations, but such shortcuts could have observable effects – like cosmic rays scattering of the computational lattice – and these effects may be needed to be corrected on the observer's level, or simulation will be "unmasked". Thus, even physical simulation needs some corrections on the observable level.

The simplest simulations, which simulate only one observer, only for a short period of time and only simple experiences require dozens of orders of magnitude less computational power than full universe, long time, high fidelity simulations. Could this mean that one is in a short and simple simulation which may be as long as just one OM, the same way as a BB?

One's simulation cannot be too simple: first, it should be able to support consciousness, and should not be full of just p-zombies and NPCs. Second, an ability to perform tasks that one typical does should be supported. This creates a lower bound of the computational complexity of one's simulation: it is not just a game or weather simulation. (However, in the novel "t" by Pelevin a character learns that he is just a character of a novel and cannot have his own thoughts – so even a very simple simulation may still think about whatever its creator wants (Pelevin, 2009).)

*1.2.4. Cost of the simulation*

Running a simulation has large computational costs. For example, simulating every atom in the whole observable universe may require more than $10^{80}$ flops if the simulation runs with the same speed as a real life, based on the number of elemental particles in the universe, which is around $10^{80}$ and high rate of collisions of atoms in stars [ref]. On the other end, simulating just one human mind may be done for around $10^{15}$-$10^{20}$ flops (Grace, 2015). It is important to emphasize that from inside the simulation one does not know what resources are available and if these numbers mentioned are trivial or not.

This suggests that computationally less expensive simulations should numerically dominate. However, less expensive simulations may have weaker internal control over glitches and bugs.

*1.2.5. Relation of the simulated world to reality*

A simulation could be either a model of the real world, or complete illusion full of absurd things. This could be represented as three main types of relation between simulation and reality:
- Ancestral simulations of a past civilization
- Possible civilization simulation
- Impossible worlds (fantasy)

*1.2.6. Types of nested simulations*

A simulation of an advanced civilization could also run its own simulations. Greene note that this will quickly result in a computational explosion which implies high probability of termination (Greene, 2018). Komosinski wrote that even current virtual machines use a few levels of nested simulation of the computer environment (Komosinski, 2018).

- One level
- Two levels
- Multilevel nested simulation.
- Uncertain for all participants, in style of classical Chinese philosopher question about reality of a dream: "Once upon a time, I, Chuang Chou, dreamt I was a butterfly, fluttering hither and thither, a veritable butterfly, enjoying itself to the full of its bent, and not knowing it was Chuang Chou. Suddenly I awoke, and came to myself, the veritable Chuang Chou. Now I do not know whether it was then I dreamt I was a butterfly, or whether I am now a butterfly dreaming I am a man"(Soothill, 1923).

Another property of multilevel ancestral simulations is that many of them likely end in the ~21st century. If a 24th-century-civilization simulates a 23rd-century-civilization, it should include all 22nd-century-civilization simulations that are run by the 23rd-century-simulation, and so on. But it all ends near the current time, as humanity cannot yet run full simulations.

Also, any level down may be numerically larger: a $23^{rd}$-century simulation could run thousands of $22^{nd}$-century simulations, and each of them could run thousands of $21^{st}$-century simulations, so at the end, any $23^{rd}$-century simulation could include millions of $21^{st}$-century simulations. Thus, it should not be surprising that we find ourselves in the $21^{st}$-century, but not later.

*1.2.7. Types of creators of the simulations*

A simulation could have different type of hosts of which superintelligent AIs seems most probable now, but guessing exact type of such AI may be beyond our epistemic capabilities
- Humans: human brain in the state of dreaming, daydreaming, storytelling, hallucinating or split personality; art objects: books, movies, games
- Superintelligent AIs: Young AGI in early stages of its development; galactic size AGI; Unfriendly evil AGI
- Natural processes, like BBs.
- Aliens: human-like, completely non-human, alien AI
- God: transcendental (beyond our ability to comprehend) or mathematical
- Simulated being in a multilevel simulation

*1.2.8. Special types of simulations*

There are also other variants of simulations, including:
- Simulations with an afterlife (after death, a character moves into another simulation): it is not exactly the same as resurrectional simulation, as it provides immortality to the beings which never existed in any reality just on moral grounds.
- Abandoned simulations: they are still running, but their creator lost interest in them
- Terminating simulations: simulations which are in the middle of the process of shutting down, which could be slow in subjective time.
- Simulations with miracles ("Miracles" here are observations which strongly deny laws of physics as humanity knows them and result in a large level of surprise; in Schmidthuber terms, they have a high level of novelty (Schmidhuber, 2010)).

*1.2.9. Observer in a simulation*

The observer could be in different locations:
- Observer is simulated by the same engine as the whole simulation
- Observer is outside of the simulation, like a gamer is outside of the game
- Brain in a vat: actual biological brain is fooled by illusion created by manipulating its senses (some VR art could be of this category)

*1.2.10. Consciousness in simulation*

At least some simulated minds may not have subjective experiences in the sense of having qualia. They could be NPC which produce only simple preprogrammed behavior. Or they could be full-blown p-zombies.

If one found oneself only as a conscious mind, and if most simulations use only p-zombies, it would be a serious argument against Bostrom's SA, because despite numerical abundance of simulations compared with the real world, the fraction of simulation with consciousness could be really small and it will not outweigh real worlds.

*1.2.11. Suffering in simulations*

Some claim that running any past-simulation would be a mindcrime, as there will be suffering humans. Based on this, they reject the existence of past simulations, as future AGI will, they hope, be benevolent.

However, counter-selection works here: The world has much unimaginable suffering of dying children, e.g. cancer patients. As most benevolent AIs will not create past simulations with human suffering, it means that our host – if any exists – is not a benevolent AGI and is more or less indifferent to human suffering. In other words, if Benevolent AIs will not run simulations, this will shift the proportion of simulations to the evil and indifferent hosts. This would still reduce the probability that we are in a simulation at all if fewer AGIs run simulations, but if the expected number of evil AGI's simulation is larger than number of real worlds, the difference will be still in favor of SA, perhaps shifting from 99.999 to 99.9 percent.

On the other hand, even a benevolent AGI may run past simulations, if the AGI turns off experiences of beings during the periods of intense pain. Such simulated beings could still have bleak and false memories about the periods of intense pain, but during the moment of the pain above certain threshold, they could be turned into p-zombies.

*1.2.12. Material simulations*

An advanced civilization could even create "material simulations", which would consist of real objects and people, but the people living in them would have an incorrect understanding about the model of the world and who is controlling it. Current examples of "material simulations" are role-playing games and a fictional example is the movie "Truman Show". Another well-known fictional example of material simulation is the TV series Westworld; a real-world example is theatre. It was speculated that alien civilizations could create Space Zoos and humanity could live in one of them (Webb, 2015): aliens control humans' lives but prefer to be hidden. The SA is applicable to the "space zoo" too: a galactic size civilization could create millions of habitable worlds.

*1.2.13. Boltzmannian simulations, hostless and abandoned simulations*

Armstrong (Armstrong, 2018) suggested that most observers should find themselves not in the real worlds, and will not be Boltzmann brains (BB) appearing out of vacuum fluctuations, but will be created by a computer, which itself is a Boltzmann brain. Such simulation without a simulator could be called hostless simulation. There could be some other type of it, including possible simulations in mathematical universe, or even pure chains of observer-moments, like in Egan's dust theory (Egan, 2009) formalized by Mueller (Mueller, 2017). This remains very speculative, so we cannot estimate the probability of being in BBs.

Also, there could be simulations which are still running but are abandoned by their host.

The general feature of all hostless simulations seems to be that there will be much more glitches, miracles and slow termination events than in the carefully managed simulations.

*1.3. Combining different conclusions about the most probable type of simulation in which one may be located*

Based on all the above, one could estimate what is the most probable type of simulation in which humanity is currently living. Obviously, this is a very tentative suggestion, as living inside a simulation makes it very difficult to model correctly the outside world and especially predict actions of an ensemble of superintelligences.

The most probable simulation type is:

- Observer-centered.
- Inexpensive.
- Scientific or entertainment simulation of a possible world history near the singularity.
- A simulation specially designed to get an advantage in measure, perhaps a resurrectional simulation.

**Part 2. Simulation's termination risks**

*2.1. Overview of the risk of the termination of a simulation*

Bostrom listed the risk of the simulation's termination as one of the serious existential risks. As short simulations are more probable than long whole universe simulations, the end of simulation could be nigh.

Why the simulation may be terminated, listed from more probable to less probable according to our understanding:

1) Overloading of the computational resources of the base reality. The computational resources needed to run the simulation of humanity could exceed available resources. For example, running computationally intensive experiments on Earth may overstretch the simulation of humanity if the simulation were generally not doing high resolution for most processes (e.g. it is not simulating atoms). For example, the bitcoin network consumes now, in 2019, around $10^{23}$ flops, and it cannot be easily approximated by a simulator. As Greene wrote: "It is partly for this reason that ancestor simulations entail a termination risk to those that create them" (Greene, 2018) because those that create them could be in a simulation. This could be an objection to the SA, as any civilization may be afraid to overload its simulator's base-level computation capability, and thus not run simulations. However, this objection does not work as it is (a) based on the assumption that the SA is true and there is another simulation level and (b) it is not universal and at least some civilizations may accept some termination risk. Greene wrote: "If we were to program our ancestor simulations to terminate before the simulated inhabitants switch on their own ancestor simulations, then that would provide us with evidence that our own simulation would be similarly programmed if we inhabit one."

2) Singularity simulation terminates shortly after AGI creation. Such a simulation will have a computational complexity crisis shortly after the singularity, as the newborn superintelligent AI will likely create its own singularity simulations etc. This will hyper exponentially increase the demand for the computation power on the base reality level. Thus, singularity simulation will be shut down shortly after the singularity, but if it produced a variant of benevolent AGI, the AGI could merge with its creator. Evil AGI would be turned off or punished, depending on how the "acausal war" between different possible AGIs unfolds.

This means that when humanity is close to creating an AGI, humanity will need to make a choice about Greene's "decision-theoretical uncertainty" conundrum: if humanity allows such AGI to create simulations and terminate them when they become nested, humanity is most likely in such a simulation and will be soon terminated.

Note that Greene's "decision-theoretical uncertainty" conundrum works best only in the case of classical SA with ancestral simulations, but loses its power in the case of the Universal SA, which assumes that humanity is most likely simulated by completely non-human beings. In that case, one can still use something like functional decision theory (Yudkowsky & Soares, 2017) to argue that humanity's unknown creators will use the same decision logic as humanity about the decision not to run or terminate the simulation. However, in the infinite universe there will be non-human and risk-taking actors who may still decide to run the simulation.

If such AGI runs resurrectional simulations and puts people in paradise-like afterlives after their simulation terminates, simulation termination would be less of a problem for humanity. Thus, humanity should allow simulations of past minds only on the condition that these minds will not be terminated after their simulation is terminated, but as Greene said, uploaded into paradise (Greene, 2018). Running very long afterlife simulations for many simulated people will still be computationally difficult, but less so than the complexity explosion in a nested simulation. Also, eventually minds could merge with each other resulting in less computationally expensive world. Furthermore, the afterlife could be run slowly to reduce computational power requirements.

3) Simulation of global catastrophic risks. The simulation could be a Fermi simulation designed to test different theories about the end of the world. The initial conditions of the Fermi simulation are those that are needed to test if a certain type of civilization will have a global catastrophe, so the catastrophe is highly probable. For example, it could test how a nuclear race evolves in bipolar world and how often this results in a catastrophic nuclear war, from which the civilization cannot recover.

4) Glitches or viruses. These could accumulate to a level where the simulation needs to be reloaded. If a simulation has been run very large number of times by an attentive host, most bugs are cleaned on early stages. Only in hostless or abandoned simulation glitches and viruses will accumulate.

5) Simulation awareness. The beings inside the simulation start to realize that they are in a simulation, so it is not a simulation anymore and cannot be used as such. Pure philosophizing about being in a simulation would likely not be enough to cause termination, as such theorizing it is likely inevitable. However, observing some undeniable glitches would destroy the purpose of the simulation as an illusion of reality, unless the simulation is rewound and corrected.

6) Game over. The simulation finally solves the unknown-to-humanity task, and there is no need to run the simulation any more. Most tasks do not require a simulation to run until the heat death of the universe, so simulations would typically have shorter duration than real life. Even movies, games and dreams typically have short duration compared with the duration of human life. However, some simulations may run very long in internal life, like the Civilization game.

7) Accidental termination. Some events in the "real world" terminate the simulation. A person could wake from a dream if her alarm rings. A computer could experience a power outage.

8) Termination accumulation the nested simulation. Higher-level simulation (closer to the actual reality) in nested simulations turns off. Interestingly, nested simulations are not dominant in our world currently, i.e., one is not often dreaming about seeing a movie in which one is writing a book about a computer gamer. Most human-created simulations are one level, and it is exactly what one should expect from the idea of domination of computationally inexpensive simulations. Multilevel simulation is inherently complex. It means that the possible simulation of humanity would be likely terminated before it would be capable of generating many multilevel simulations, that is shortly after the appearance of superintelligent AI, as was mentioned by Bostrom and Greene (Bostrom, 2003; Greene, 2018).

9) Natural end. A simulation will probably be terminated after all beings in it have died or all natural processes have stopped.

10) Unknown threshold. The simulation of humanity could pass an unknown threshold after which it must be terminated. As humanity is now alive such threshold, if it exists, is ahead of us. The threshold could not only be AGI creation, but some moral threshold, such as a large population, intense suffering, or sentient computers. Alternatively, the threshold could be technological, such as self-replicating nanotechnology or genetically engineered babies or philosophical, like the capability of thinking about being in a simulation clearly. Given the accelerated rate of progress, such thresholds will probably happen almost simultaneously in historical time, probably somewhere in the 21st century. Other thresholds could be more remote, like star-travel or destroying life on Mars (as it could be taken as an evidence of our hostile nature) or colonizing the galaxy.

Robin Hanson recommended acting in a way that will make one's simulation more interesting to the possible hosts thus lowering chances of the simulation termination: "If you might be living in a simulation then all else equal you should care less about others, live more for today, make your world look more likely to become rich, expect to and try more to participate in pivotal events, be more entertaining and praiseworthy, and keep the famous people around you happier and more interested in you" (Hanson, 2001).

On the other hand, even discussing simulations maybe an informational hazard as it could increase chances that the simulation will be terminated. However, as the topic has been discussed for ~20 years without the simulation's termination, this is evidence for the host's tolerance to the simulation awareness. This evidence maybe rather weak given observational selection effects: even if most self-aware simulations were terminated, we could find ourselves only where the termination threshold is higher than zero-tolerance. However, it may not be possible to simulate the 21st century without simulating some scientists who think that they may be in a simulation. Hanson wrote that "simulations... tend to be ended when enough people in them become confident enough that they live in a simulation" (Hanson, 2001). It appears that Bostrom did not worry about his SA possibly causing simulation shutdown, and thus being the ultimate hazardous information, though he wrote a lot about information hazards in other domains (Bostrom, 2011).

From all listed above ideas about the termination of the simulation, the idea of the termination because of the computational overload appears to be the most probable, as it inevitably follows from the setup of the SA. This may coincide with the fulfilment of the goal of the simulation, if the simulation is created to explore different types of the singularity and/or global catastrophic risks near it. If the simulation finishes before self-improving AGI, most interesting data will be not learned, but if it finishes after, its computational price will grow enormously.

Termination because of observation of glitches seems to be much less probable, as hosts could create glitch-less simulations if they prefer and polish it in many runs or rewind and correct.

## 2.2. The end of a multilevel simulation

The idea of a multilevel simulation was discussed by (Komosinski, 2018) and (Torres, 2014). As we discussed above, lower level simulations are also more numerically abundant, but computationally simpler and, in case of ancestor simulations, are simulating earlier periods of history.

In a nested simulation, all risks of termination are accumulated, as was mentioned by Torres (Torres, 2014). If there is an infinite chain of nested simulations, one of them, as Torres said, is surely now turning off. As we are not turning off now, it seems evidence that humanity is not in an infinite chain, or even sufficiently large chain of simulation.

However, for a nested simulation, it is typical to have time acceleration: time in simulated world runs more quickly than in the hosts' worlds. Time acceleration multiplies over nested simulations, $t^n$, where n is the number of simulation levels and t is a typical time acceleration parameter. (For human-created simulations, it is typical to have time acceleration of a few orders of magnitude: in a book, years of the internal time will pass, while one could read it in 1 night, which imply 1000 times dilution). Time acceleration in a nested simulation is an exponential process $t^n$, and it may outperform accumulating of the termination probability.

Also, any termination is not an instant process: some amounts of the computations are happening inside different parts of a computer a few milliseconds after the power is switched off [ref]. Because of the time acceleration, such milliseconds could be translated into years in subjective time of a simulation, especially in the one which is a few levels downstream.

For an observer inside a simulation, such slow termination could look like a slow and maybe painful process. Different servers could turn off with different speeds, so different parts of the world may stop existing at different rates. Maybe, at first, stars will disappear.

Such a terminating simulation is a subtype of abandoned simulations, and glitches are much more probable in it.

In a nested simulation, errors could accumulate from top down, making the lower-level simulation more glitchy. However, ethical considerations (like suffering prevention) should grow downstream, as higher-level hosts could intervene if lower-level ones are making bad outcomes in their simulations. But time acceleration effect will make observations of such actions very rare for downstream simulations.

Lower level simulations should be also computationally simple: as all actual calculations are happening on the highest real level, each level could provide only a part of its computational resources for running sub-simulations. Thus, at the lowest level, there may be many 21st century simulations, running with the highest economy of computational resources, which increase chances of glitches.

In a multilevel simulation, each level will not be certain about its own reality. Also, exactly the same simulation could be run by different hosts, creating indexical uncertainty.

## 2.3. A global catastrophe in the simulation

A simulation which is modeling a global catastrophe is not the same that a terminating simulation from formal point of view, but doesn't have much difference for the simulation inhabitants. There are several overlapping reasons why we could be located in such type of simulation:

1) Fermi simulations created by alien AIs to solve the Fermi paradox are designed in the way that they should model the world in the vicinity of a global catastrophe.
2) Singularity simulations, or models of the first AI appearing, are modeling the period of history when different global risks peak. The period before creating of AGI is also a period of quick development of other potentially dangerous technologies, like synthetic biology (Millett & Snyder-Beattie, 2017; Turchin, Green, & Denkenberger, 2017), colliders (Kent, 2004), advance nuclear weapons etc. The events before singularity could quickly accelerate, driven by arms races and different instabilities. Singularity simulations may also end up with unstable or evil AI which will cause an observable catastrophe, like a paperclipper.
3) Simulations as computer games centered around attempts to save the world.
4) Second leg of the Bostrom's argument: if we are not in the simulation, it is best explained by the fact that there is some universal reason for the civilizations' extinction.

Obviously, alien AGI will be also interested in outcomes of the attempts of AGI creation of different planets, so there is no practical difference between Fermi simulations by alien AGI and ancestral Singularity simulations by our own superintelligent AI.

### *2.4. Doomsday Argument for simulations predicts termination soon*

The DA and SA do not cancel each other, as was assumed by Aranyosi (Aranyosi, 2004), but are actually supporting. Combined, they claim: "A conscious observer is most likely in a simulation, and it will probably end soon". The reason for this is that the DA logic should be applied only to the duration of the simulation, but not to the number of observers in all real and simulated worlds, as observer birth rank is known and ordered only inside any given simulation. While the birth order in the simulation could be faked by creators, it make the DA inside the simulation even stronger: if the simulation was created – with all false memories – just yesterday, it will most likely will end tomorrow.

Thus, Doomsday argument for simulations runs as following: using the Copernican mediocrity principle (Gott, 1999), an observer can suggest that the observer is located somewhere in the middle of the duration of his-her simulation, as his-her moment now is randomly chosen from the whole duration of the simulation. This means that if the median total duration of the simulation duration is T, when it will likely end in T/2 from now.

Perhaps one's simulation was created just yesterday with all his-her memories? It means even shorter duration of the simulation thus making DA in simulation even stronger. But to have all the memories in order, these memories should be simulated for some period of time, so the idea of the simulation with false memories appears to be self-defeating as it assumes the existence of another simulation which prepared these false memories. However, these false memories could be prepared once for many short runs, the same way as false memories were prepared for characters of the TV-series Westworld, where each short run of a character was only one day, after which she was reloaded with the same old memories.

Anyway, T cannot be very short, or much more complex structure of the nested simulation is needed to describe appearance of the false memories. For Fermi simulations, it is probably reasonable to simulate human history between the beginning of the 20th century and the Singularity, which is approximately 150 years, as during this period most critical choices were made.

Moreover, simulation argument itself predicts shorter simulations. As we discuss above, simpler simulations should numerically dominate. This also means that shorter duration of internal time simulations also should numerically dominate: that is, the simulations which simulate only one human's life are more numerous than simulations which simulate whole the human history for 5000 years, and billion years old simulations are extremely rare.

Lewis also compared SA and DA (Lewis, 2013), and finds that SA works, but DA doesn't. However, he did not analyze the simple Gott's DA, and instead used the Sleeping Beauty model for his analysis, and did not apply the DA to an observer inside the simulation.

### *2.5. Immortality in simulations*

Existence of an extremely large number of simulations created by many different civilizations in a potentially infinite universe implies that there are other copies of each person in different simulations. There is a very large but finite number of possible people, limited by different combinations of atoms, and if the number of the simulations is larger than the number of possible people, people will repeat in the simulations.

Moreover, if humanity is located in an ancestral simulation, it will likely be run many times with small variations (according to the assumptions of SA), and thus many of its elements will be repeated exactly or with small variations, including human beings.

Thus, if one's simulation were turned off, other simulations with other copies would continue to exist and this could be a form of immortality. For example, if there are 10 exact copies of oneself in different simulations, and 9 of the simulations are terminated, one copy will survive, which may or may not be interpreted as some form of immortality depending on how one treats personal identity and measure, see (Turchin, 2018a).

If one were an avatar of some gamer, the end of the simulation would look like the awakening for that avatar in the body of gamer, and merging of memories of the gamer and the avatar's experiences in the simulation. This is analogous to the way that one's memories about a dream merge with real life after awakening.

**Conclusion**

Humanity's location at the beginning of the 21st century could be best explained by the fact that this period is in a scientific Fermi simulation by an alien civilization or future humanity-based AGI simulating variants of its own origin, which could be called a "singularity simulation". This means that humanity could be tested for different scenarios of global catastrophic risks, and no matter what the result of the test is, the simulation of us would be turned off relatively soon, in tens to hundreds of years from now.

Also, humanity could be in a gaming simulation of some posthuman or alien beings, which is similar to a Fermi simulation but created for recreational purposes. This has even worse termination risks because it could be ended for arbitrary reasons (such as boredom).

An alternative explanation is that humanity is located in a resurrectional simulation which is optimized to increase the measure of observers in it (that is, the probability that a random observer will be in such simulation). Such simulation would be created by benevolent AGI and does not have termination risks because all beings in such a simulation are moved into an afterlife simulation after death and later are adapted to live in the real world (Turchin, 2019). Such resurrectional simulation will eventually terminate, but all minds from it will be saved.

**Literature**

Alexander, S. (2018). A type of simulation which some experimental evidence suggests we don't live in. ArXiv Preprint ArXiv:1808.03225.

Almond, P. (2006). Many-Worlds Assisted Mind Uploading: A Thought Experiment. Retrieved from https://web.archive.org/web/20110513092111/http://www.paul-almond.com/ManyWorldsAssistedMindUploading.htm

Aranyosi, I. A. (2004). The Doomsday Simulation Argument. Or why isn't the end nigh and you're not living in a simulation.

Armstrong, S. (2018). Are you in a Boltzmann simulation? - LessWrong 2.0. Retrieved February 4, 2019, from LessWrong website: https://www.lesswrong.com/posts/ygELzNSAF5nzLXD7j/are-you-in-a-boltzmann-simulation

Armstrong, Stuart. (2010). The AI in a box boxes you - LessWrong 2.0. Retrieved April 16, 2019, from LessWrong website: https://www.lesswrong.com/posts/c5GHf2kMGhA4Tsj4g/the-ai-in-a-box-boxes-you

Beane, S. R., Davoudi, Z., & Savage, M. J. (2014). Constraints on the Universe as a Numerical Simulation. The European Physical Journal A, 50(9), 148.

Bennett, C. H. (2003). Notes on Landauer's principle, reversible computation, and Maxwell's Demon. Studies In History and Philosophy of Science Part B: Studies In History and Philosophy of Modern Physics, 34(3), 501–510.

Borges, J. L. (2000). The library of Babel. David R. Godine Boston, MA.

Bostrom, N. (2002a). Existential risks: Analyzing Human Extinction Scenarios and Related Hazards. Journal of Evolution and Technology, Vol. 9, No. 1 (2002).

Bostrom, N. (2002b). Self-locating belief in big worlds: Cosmology's missing link to observation. The Journal of Philosophy, 99(12), 607–623.

Bostrom, N. (2003). Are You Living In a Computer Simulation? Published in Philosophical Quarterly (2003) Vol. 53, No. 211, Pp. 243-255.

Bostrom, N. (2013). Anthropic bias: Observation selection effects in science and philosophy. Routledge.

Bostrom, N. (2014). Superintelligence. Oxford: Oxford University Press.

Bostrom, Nick. (2011). Information hazards: a typology of potential harms from knowledge. Review of Contemporary Philosophy, (10), 44–79.

Chalmers, D. (1996). The Conscious Mind. Oxford University Press, New York.

Christiano, P. (2011). The Absolute Self-Selection Assumption - LessWrong 2.0. Retrieved April 15, 2019, from LessWrong website: https://www.lesswrong.com/posts/QmWNbCRMgRBcMK6RK/the-absolute-self-selection-assumption

Christiano, P. (2012). A formalization of indirect normativity. Retrieved April 16, 2019, from Ordinary ideas website: https://ordinaryideas.wordpress.com/2012/04/21/indirect-normativity-write-up/

Dainton, B. (2002). Simulation Scenarios: Prospects and Consequences. Retrieved from https://philpapers.org/archive/DAIILS.pdf

Egan, G. (2009). Dust Theory FAQ. Retrieved from http://www.gregegan.net/PERMUTATION/FAQ/FAQ.html

Gackenbach, J., & LaBarge, S. (2012). Conscious mind, sleeping brain: Perspectives on lucid dreaming. Springer Science & Business Media.

Gerig, A., Olum, K. D., & Vilenkin, A. (2013). Universal doomsday: analyzing our prospects for survival. Journal of Cosmology and Astroparticle Physics, 2013(05), 013.

Good, I. J. (1965). Speculations concerning the rst ultraintelligent machine. In F. l. Alt & M. rubino (Eds.), Advances in.

Gott, J. R. (1999). The Copernican Principle and Human Survivability. Human Survivability in the 21th Century. Transactions of the Royal Society of Canada, Series VI, 9, 131–147.

Grace, K. (2015). Brain performance in FLOPS – AI Impacts. Retrieved January 23, 2018, from https://aiimpacts.org/brain-performance-in-flops/

Greene, P. (2018). The Termination Risks of Simulation Science. Erkenntnis, 1–21.

Hanson, R. (2001). How to live in a simulation. Journal of Evolution and Technology, 7(1).

Hutter, M. (2000). A theory of universal artificial intelligence based on algorithmic complexity. ArXiv Preprint Cs/0004001.

Kent, A. (2004). A critical look at risk assessments for global catastrophes. Risk Analysis, 24(1), 157–68.

Komosinski, M. (2018). Universes and simulations: Civilizational development in nested embedding. Foundations of Computing and Decision Sciences, 43(3), 181–205. https://doi.org/10.1515/fcds-2018-0010

Kurzweil, R. (2006). Singularity is Near. Viking.

Lewis, P. J. (2013). The doomsday argument and the simulation argument. Synthese, 190(18), 4009–4022.

Medvedev, D. (2003). Are We Living In Nick Bostrom's Speculation | Reason | Reality. Retrieved from https://www.scribd.com/document/155089869/Are-We-Living-In-Nick-Bostrom-s-Speculation

Millett, P., & Snyder-Beattie, A. (2017). Human Agency and Global Catastrophic Biorisks. Health Security, 15(4), 335–336.

Mueller, M. P. (2017). Law without law: from observer states to physics via algorithmic information theory. ArXiv:1712.01826 [Physics, Physics:Quant-Ph]. Retrieved from http://arxiv.org/abs/1712.01826

Pelevin, V. P. (2009). T. Moskva: M.:"Eksmo".

Sandberg, A. (2016). Energetics of the brain and AI. ArXiv Preprint ArXiv:1602.04019.

Schmidhuber, J. (2010). Formal Theory of Creativity and Fun and Intrinsic Motivation Explains Science, Art, Music, Humor. Retrieved April 22, 2019, from http://people.idsia.ch/~juergen/creativity.html

Soothill, W. E. (1923). The three religions of China: Lectures delivered at Oxford. Oxford University Press.

Sotala, K., & Gloor, L. (2017). Superintelligence as a Cause or Cure for Risks of Astronomical Suffering.

Tegmark, M. (1998). The Interpretation of Quantum Mechanics: Many Worlds or Many Words? Fortschritte Der Physik, 46(6-8), 855–862. https://doi.org/10.1002/(SICI)1521-3978(199811)46:6/8<855::AID-PROP855>3.0.CO;2-Q

Tegmark, M. (2009). The multiverse hierarchy. ArXiv Preprint ArXiv:0905.1283.

TheTripleAffirmative. (2019). The Cacophony Hypothesis: Simulation (If It is Possible At All) Cannot Call New Consciousnesses Into Existence - LessWrong 2.0. Retrieved April 15, 2019, from LessWrong website: https://www.lesswrong.com/posts/sMjMYNWMKjrrN35pX/the-cacophony-hypothesis-simulation-if-it-is-possible-at-all

Tipler, F. J. (1997). The physics of immortality: Modern cosmology, God, and the resurrection.

Torres, P. (2014). Why Running Simulations May Mean the End is Near. Retrieved from https://ieet.org/index.php/IEET2/more/torres20141103

Turchin, A. (2017). Messaging future AI. Manuscript. Retrieved from https://goo.gl/YArqki

Turchin, A. (2018a). Forever and Again: Necessary Conditions for the "Quantum Immortality" and its Practical Implications.

Turchin, A. (2018b). Preventing s-risks via indexical uncertainty, acausal trade and domination in the multiverse - LessWrong 2.0. Retrieved April 16, 2019, from LessWrong website: https://www.lesswrong.com/posts/eWx8LzasNutvmkPMf/preventing-s-risks-via-indexical-uncertainty-acausal-trade

Turchin, A. (2019). You only live twice: resurrection of the dead in computer simulations.

Turchin, A., & Chernyakov, M. (2018). Classification of Approaches to Technological Resurrection. Under Review in Post-Humans Studies. Retrieved from https://philpapers.org/rec/TURCOA-3

Turchin, A., & Denkenberger, D. (2017). Levels of self-improvement. Manuscript.

Turchin, A., & Denkenberger, D. (2019). Classfication of ETI riks. Inder Review in JBIS. Retrieved from https://philpapers.org/rec/TURGCR

Turchin, A., Green, B., & Denkenberger, D. (2017). Multiple Simultaneous Pandemics as Most Dangerous Global Catastrophic Risk Connected with Bioweapons and Synthetic Biology. Under Review in Health Security.

Volpicelli, G. (2016). This Transhumanist Records Everything Around Him So His Mind Will Live Forever. Vice.Motherboard. Retrieved from https://motherboard.vice.com/en_us/article/4xangw/this-transhumanist-records-everything-around-him-so-his-mind-will-live-forever

Wall, M. (2018). We're Probably Living in a Simulation, Elon Musk Says. Retrieved April 17, 2019, from Space.com website: https://www.space.com/41749-elon-musk-living-in-simulation-rogan-podcast.html

Webb, S. (2015). If the Universe Is Teeming with Aliens... Where is Everybody?: Seventy-Five Solutions to the Fermi Paradox and the Problem of Extraterrestrial Life. Springer.

Whitworth, B. (2008). The physical world as a virtual reality. ArXiv Preprint ArXiv:0801.0337.

Yampolskiy, R. (2016). On the Origin of Samples: Attribution of Output to a Particular Algorithm. ArXiv Preprint ArXiv:1608.06172.

Yampolskiy, R. (2018). Minimum Viable Human Population with Intelligent Interventions. ResearchGate. http://dx.doi.org/10.13140/RG.2.2.33337.83042

Yampolskiy, R. (2019). It is only for a few years right before AGI is... -. Retrieved April 17, 2019, from Facebook post website: https://www.facebook.com/photo.php?fbid=10216367888522126&set=a.2227672163627&type=3&theater

Yampolskiy, R. V. (2017). Detecting Qualia in Natural and Artificial Agents. ArXiv:1712.04020 [Cs]. Retrieved from http://arxiv.org/abs/1712.04020

Yampolskiy, R. V. (2019). Personal Universes: A Solution to the Multi-Agent Value Alignment Problem. ArXiv:1901.01851 [Cs]. Retrieved from http://arxiv.org/abs/1901.01851

Yudkowsky, E. (2008). Where Physics Meets Experience - Less Wrong. Retrieved February 8, 2018, from LessWrong website: http://lesswrong.com/lw/ps/where_physics_meets_experience/

Yudkowsky, Eliezer, & Soares, N. (2017). Functional Decision Theory: A New Theory of Instrumental Rationality. ArXiv:1710.05060 [Cs]. Retrieved from http://arxiv.org/abs/1710.05060